\documentclass[aps, prx, twocolumn, superscriptaddress, ]{revtex4-1}
\usepackage{amsmath,amssymb,amsfonts,amsbsy}
\usepackage{graphicx}
\usepackage{amsmath}
\usepackage{subfig}
\usepackage{float}
\graphicspath{{./figures/}}
\usepackage{dcolumn}
\usepackage{soul}
\usepackage{bm}
\usepackage{comment}
\usepackage{caption}
\usepackage{subfig}
\usepackage{stmaryrd}
\usepackage{multirow}
\usepackage{mathtools}
\usepackage{array}
\usepackage{amsmath}
\usepackage{color}
\usepackage[normalem]{ulem}
\usepackage[per-mode=symbol]{siunitx}
\usepackage{upgreek}

\usepackage[nodisplayskipstretch]{setspace}
\usepackage[labelfont=bf]{caption}
\usepackage[figurename=Fig.]{caption}

\DeclareSIUnit \belm {Bm}

\usepackage[colorlinks=true,breaklinks=true,allcolors=blue]{hyperref} 
\usepackage[capitalise,nameinlink]{cleveref} 
\usepackage{booktabs}
\usepackage[utf8]{inputenc}
\usepackage[T1]{fontenc}

\newcommand{\app}[1]{\hyperref[app:#1]{Appendix~\ref*{app:#1}}}

\def\@setaltaffiliation{\vspace{-\baselineskip}\def\altaffiliation##1{\@par##1\@addpunct.}\altaffiliationes}
\def\@setaltaffiliation{\vspace{-\baselineskip}\def\altaffiliation##1{\@par##1\@addpunct.}\altaffiliationes}
\RequirePackage{lineno}
\setpagewiselinenumbers

\begin{document}
\title{Hardware-aware Training Techniques for Improving Robustness of Ex-Situ Neural Network Transfer onto Passive TiO$_{\textrm{2}}$ ReRAM Crossbars}
\author{
Philippe~Drolet$^{1,2}$, Raphaël~Dawant$^{1,2}$, Victor~Yon$^{1,2}$, Pierre-Antoine~Mouny$^{1,2}$, Matthieu~Valdenaire$^{1,2}$, Javier~Arias~Zapata$^{1,2}$, Pierre~Gliech$^{1,2}$, Sean~U.~N.~Wood$^{4}$, Serge~Ecoffey$^{1,2}$, Fabien~Alibart$^{1,2,3}$, Yann~Beilliard$^{1,2}$, Dominique~Drouin$^{1,2}$
\\\footnotesize$^1$\textit{3IT (Institut Interdisciplinaire d’Innovation Technologique), 3000 Boulevard de l’université, Sherbrooke, QC, Canada}
\\\footnotesize$^2$\textit{Laboratoire Nanotechnologies et Nanosystèmes, IRL-LN2, CNRS, Université de Sherbrooke}
\\\footnotesize$^3$\textit{Institute of Electronics, Microelectronics and Nanotechnology (IEMN), Université de Lille, 59650 Villeneuve d’Ascq, France}
     \\\footnotesize$^4$\textit{Department of Electrical and
Computer Engineering
Université de Sherbrooke
Sherbrooke, QC, Canada}}
\altaffiliation{Corresponding author: \\philippe.drolet3@usherbrooke.ca}
\date{\today}
\begin{abstract}
Passive resistive random access memory (ReRAM) crossbar arrays, a promising emerging technology used for analog matrix-vector multiplications, are far superior to their active (1T1R) counterparts in terms of the integration density. However, current transfers of neural network weights into the conductance state of the memory devices in the crossbar architecture are accompanied by significant losses in precision due to hardware variabilities such as sneak path currents, biasing scheme effects and conductance tuning imprecision. In this work, training approaches that adapt techniques such as dropout, the reparametrization trick and regularization to TiO$_{\textrm{2}}$ crossbar variabilities are proposed in order to generate models that are better adapted to their hardware transfers. The viability of this approach is demonstrated by comparing the outputs and precision of the proposed hardware-aware network with those of a regular fully connected network over a few thousand weight transfers using the half moons dataset in a simulation based on experimental data. For the neural network trained using the proposed hardware-aware method, 79.5\% of the test set's data points can be classified with an accuracy of 95\% or higher, while only 18.5\% of the test set's data points can be classified with this accuracy by the regularly trained neural network. 
\end{abstract}
\maketitle
\section{Introduction}
As neural networks' complex problem-solving capabilities increase, so do their energetic and computational demands. These demands have grown so much that the use of graphics processing units (GPUs) and data centers is proving to be essential for practical machine learning. Meanwhile, a plethora of applications at the edge would greatly benefit from the use of neural networks but cannot always use these deep networks due to their high energy demands \cite{edge-computing-review}. As the main limitations of neural network computations stem from the von Neumann bottleneck between the memory and processor \cite{Li2015}, in-memory computing with analog, non-volatile emerging resistive memories is among the most promising avenues of innovation that can be used to overcome this problem \cite{Amirsoleimani2020,Jeong2016,Ielmini2020}. Non-volatile memories exploit metal oxide \cite{mesoudy2021CMOS,Dalgaty2021,Kim2021,Bayat2018}, phase-change \cite{phase-change} or ferroelectric materials \cite{Nishitani2012}, among other things, to perform computations directly in memory, decreasing the need to fetch data to perform computations. In a crossbar configuration, in-memory computing relies on the fundamental Ohm's and Kirchoff's laws, which enable an energy-efficient version of the vector matrix multiplications (VMMs) at the core of neural networks. These memory devices are versatile and have proven their potential for many different types of neural networks, such as long short-term memory (LSTMs) \cite{LSTM-hardware-aware,LSTM-ReRAM}, generative adversarial networks (GAN) \cite{Chen2022} and Bayesian NNs \cite{BNNrealizationReRAM,zhou_bayesian_2021,BNNexploitingReRAM}. 
Neural networks on crossbars always require either in-situ or ex-situ learning techniques. Both of which have been demonstrated on $TiO_{2}$ crossbars \cite{Alibart2013}. The former technique trains the network online on the devices by implementing backpropagation through tuning pulses. The latter performs the training offline and thereafter the network is transferred to the crossbars by converting the obtained trained weights into conductances. The more commonly used ex-situ technique is much simpler but hardly considers hardware variabilities during training. This renders it less robust to all potential hardware non-idealities. 
\begin{figure*}[ht!]
  \centering
  \includegraphics[width= 0.9\textwidth]{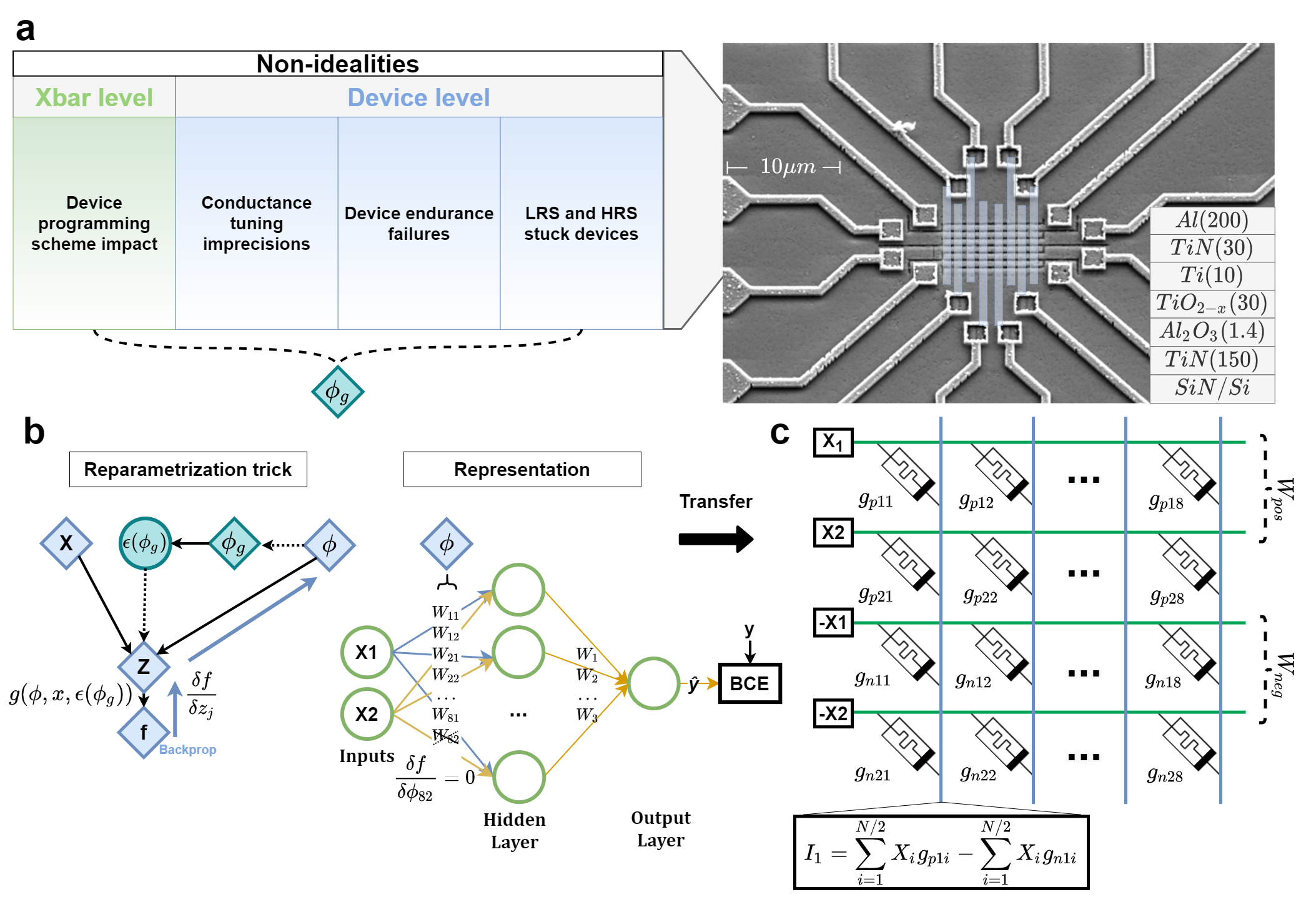}

\caption{\textbf{Schematic illustration of the main concept investigated in this work.} \textbf{a.} (Right) A scanning electron microscopy (SEM) picture of our 8$\times$8 $TiO_{2}$ crossbar architecture. The top electrodes (word lines) are visible, while the bottom electrodes (bit lines) fabricated by a damascene process have been represented in blue. The materials stack of the resistive memory devices is displayed in the inset. (Left) The different sources of variability considered during training are shown. \textbf{b.} Depiction of the training process of the neural network. The left side of the sub-figure represents the reparametrization trick in relation to the weight values. The right side is a schematic representation of the neural network investigated in the scope of a half moon toy problem.. A dotted line indicates a conversion between ranges (weight range to conductance range or vice versa). The weights are initially converted to conductances with differential weights ($G_{+} - G_{-}$). A noise parameter $\epsilon$ is then added to both the positive and negative devices through a reparametrization trick. A percentage of neurons, depicted in the schematic by a half-dashed \textit{X} in the hidden layer for weight $W_{82}$, are, similarly to dropout, deactivated by conversion to a failure state in the low resistance state (LRS) or high resistance state (HRS) (instead of by being set to 0 as they would be in the usual dropout method). The gradient on these neurons is then zeroed out during backpropagation. \textbf{c.} Physical implementation on passive crossbar array. The weights are represented by two devices to be able to represent signed values.}
\label{fig:figure 1}
\end{figure*}
 The crossbar architectures presented in the literature are often based on the 1T1R integration scheme \cite{Xue2020154A2,Yao2017}, where a transistor (1T) is cascaded with each memory (1R) to allow independent accesses to desired memories without disturbing adjacent devices. This improved control provides high programming and reading accuracies but leads to a lower integration density, higher wiring complexity and increased fabrication costs. On the other hand, passive (0T1R) crossbars exhibit excellent scalability and lower fabrication costs. In \cite{Xia2019}, a factor of 20 between the sizes of individual 1T1R (4 $\mu$m) and 0T1R ($200$ nm) memristors was reported. An average
relative conductance import accuracy of 1\% on a fully passive $TiO_{2}$ crossbar with a size of 64$\times$64 was reported \cite{Kim2021}. This crossbar had $\sim99\%$ of memristors functioning. As such, passive $TiO_{2}$ crossbars hold much promise for the future of in-memory computing with ReRAM devices.

However, these passive crossbars present multiple inherent sources of variability  \cite{varia_memristor}, which currently hinder their use at large scale. The conductance tuning imprecision that is observed when a device is programmed to a certain conductive level, unsatisfactory device yields, asymmetrical switching dynamics, the conductance drift due to relaxation effects over time and sneak path currents are a few examples of the issues afflicting this technology. These non-idealities directly affect the results of analog matrix-vector multiplications (VMM). Cumulatively, they often lead to incorrect  classifications. While some of these non-idealities can be harnessed to implement efficient stochastic neural network models in hardware \cite{Yon2022,Dalgaty2021,BNNexploitingReRAM,LSTM-hardware-aware}, they remain very detrimental for standard in-memory computing solutions and make it difficult to scale up this technology. A potential avenue that can be used to mitigate the sneak path current problems is the use of tiled crossbars; the neural network weights can be separated onto sub-crossbars called tiles. Tiles are compatible with control transistors and circuitry. These 8$\times$8 crossbar tiles offer an intermediate solution that lies between the improved controllability of 1T1R crossbars and the scalability of 0T1R crossbars. \newline

In this work, we report a novel hardware-aware ex-situ training approach that improves the robustness of neural network models against the non-idealities of $TiO_{2}$-based memory crossbars. The main contributions of this study are the introduction of new, fully data-driven and computationally simple characterization and variability modeling methods that are used jointly with a reparametrization trick during training to mitigate precision losses when weights are converted into their equivalent ReRAM conductances on a passive crossbar. This procedure is often referred to as hardware-aware training. The simple variability models do not require computationally prohibitive mathematical and physical models, and they lead to good performance without requiring multiple rounds of tuning; these models require only reproducible characterizations that are all easy to automate. Figure \ref{fig:figure 1} shows the overall hardware-aware procedure. The hardware-aware network provides accurate classification results for the half moons dataset for a much greater percentage of the simulated transfers compared to a regular neural network. This greater accuracy across transfers proves that the hardware-aware procedure is robust to significant biasing-scheme-effect conductance changes (up to 60 $\mu$S), conductance tuning imprecision and neural network weights impacted by random substitutions to account for stuck devices. The devices studied in this work are 0T1R devices, and they can be seen in Figure \ref{fig:figure 1} a and in Supplementary Figure \ref{fig:supp_sem_closeup}.

Other works, such as \cite{zhou_bayesian_2021,Strukov-hardware-aware,Mitigating-Imperfections}, employ a similar approach. In \cite{zhou_bayesian_2021}, an impressive performance was reported but the authors chose not to consider device failures and only consider 1T1R devices, therefore neglecting biasing scheme effects. Meanwhile, \cite{Strukov-hardware-aware, Mitigating-Imperfections} deal with 0T1R $TiO_{2}$ passive crossbars and implement many different variability sources such as temperature, biasing scheme effect and non-linearity in their hardware simulations and training to great effect. However, the relationship between the variability and the crossbar position of the devices is not considered in these works. The focus is instead placed on a conductance tuning technique that mitigates biasing scheme disturbances. Also, the fact that our method depends solely on empirical data means that it is also applicable to different memristive technologies with slight modifications, rendering it highly adaptable. The biasing scheme considered in these other works is the V/2 scheme; the V/3 scheme is considered in this work and the variability modeling approaches are therefore different. 

In this paper, we begin by explaining the different sources of variability that are taken into account by the hardware-aware network, along with the characterization process that led to the creation of each of these sources. We continue with a software demonstration that compares the performances of weights trained considering these variabilities and weights trained naively over 10,000 simulated ex-situ weight transfers for a simple binary classification problem. The hardware-aware network significantly outperforms the regular neural network.

\begin{figure*}[t]  
    \centering
    \includegraphics[scale = 0.20]{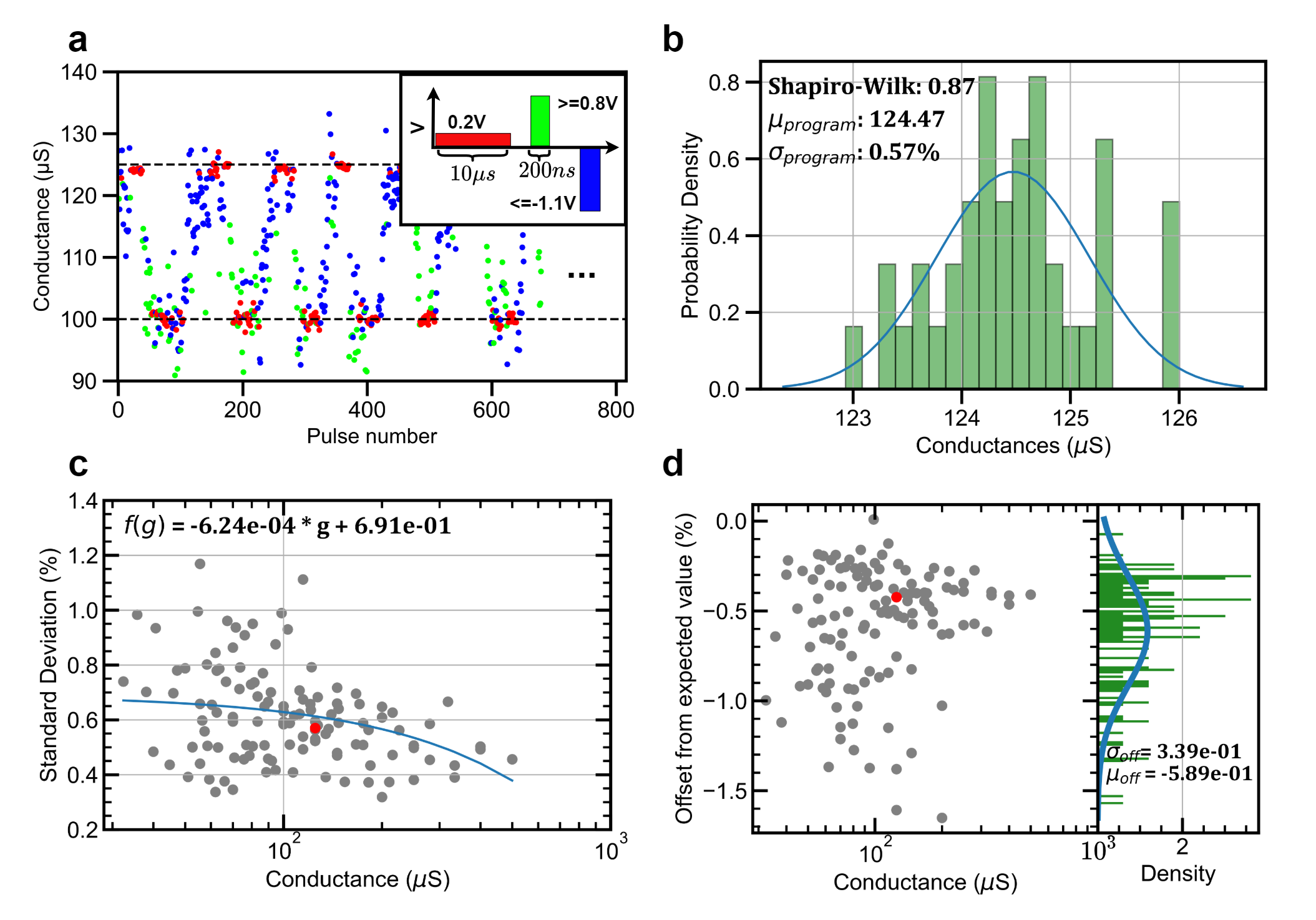}
    \caption{\textbf{Experimental measurements of the conductance tuning variability of the TiO2-based resistive memories.} \textbf{a.}\, The conductance is tuned from a low conductance state to the target conductance state using SET (in blue) and RESET (in green) pulses forty times. The last read value is then kept in memory so that  \textbf{b.}\, a normal distribution may be fitted to the programmed conductances. A conductance value of 125 $\mu$S is targeted, starting from 100 $\mu$S in this example. A simple Shapiro-Wilk test is performed to see if the null hypothesis must be rejected. A mean and a standard deviation are extracted from this distribution. These two steps were repeated eight times across the conductive range of each individual device. A total of 29 devices were tested with this protocol. \textbf{c.}\, Depiction of the relationship between all of the standard deviations of the fitted normal distributions presented in panel \textbf{b} and the conductive level of the device. The average standard deviation increases as the conductance decreases, as predicted. The fitted linear relationship between the conductance and standard deviation is expressed on the plot.  \textbf{d.}\, Depiction of the relationship of the offset with respect to the mean of the normal distributions, similar to the one shown in panel \textbf{b}, and their targeted (ideal) conductance with respect to the conductive level of the device. There is no clear correlation between the offset from the target value and the conductive level of the device. A normal distribution is instead fitted to the offsets. }
    \label{fig:conductance_tuning}
\end{figure*}  
\section{Results and discussion}

As previously mentioned, passive crossbars are afflicted by many different and often indirectly correlated non-idealities. The device-to-device and crossbar-to-crossbar variabilities are considerable and difficult to predict. Depending on a specific crossbar's properties, a transferred neural network may perform well or very poorly. The proposed approach aims at using neural networks' learning abilities to learn how to classify data points in spite of the effects that non-idealities will have on analog VMMs. These non-idealities were estimated through the extensive electrical characterization of TiO$_{\textrm{2}}$-based resistive memories fabricated in 8$\times$8 crossbar and crosspoint configurations. The considered non-idealities were those deemed to have the greatest effect on VMMs, namely the conductance tuning imprecision, biasing scheme effects and high resistance state (HRS) and low resistance state (LRS) stuck devices.

During training, the neural network weights are converted from their software range to the predetermined mean achievable conductance range $G_{R}$ for every batch of data. The weights are then shifted according to randomized variability samples using a reparametrization trick \cite{https://doi.org/10.48550/arxiv.1312.6114} integrated into a modified version of the PyTorch library \cite{NEURIPS2019_9015}.  The use of the reparametrization trick allows the Monte Carlo estimate of the expectation (every time we sample new weights) to be differentiable with respect to the weight parameters and enables backpropagation, as depicted in Figure \ref{fig:figure 1}b. The process is explained by the flowchart shown in Figure \ref{fig:figure 1} and the mathematical details are given in Supplementary Note 1. All considered non-idealities are numerically added into a matrix $\phi_{g}$ created from the conversion of the network's parameters $\phi$ to the conductance range $G_{R}$ and share the same dimensions. This matrix is converted back into the weight range after variabilities are added. The now-converted matrix is subtracted from the original parameter matrix $\phi$ to obtain the $\epsilon(\phi_{g})$ matrix used in the reparametrization trick. $G_{R}$ was determined  to be from 100 $\mu$S to 400 $\mu$S by electrical characterizations made on our 8$\times$8 crossbars \cite{mesoudy2021CMOS}. The randomized variability samples depend on the conductance tuning imprecision and the biasing scheme effects recorded in a database. Percentages of neural network weights \textit{x} and \textit{y} corresponding to LRS and HRS stuck devices are also replaced by equivalent high or low conductance values. This effectively simulates the result of an ex-situ transfer on a crossbar, as depicted in Figure \ref{fig:figure 1}c, with randomized programming failures for every batch of training data. It also ensures that the network considers randomized potential sources of hardware-related errors. The trained network will converge towards the solution that is most robust to the average of all potential crossbars, considering our TiO$_{\textrm{2}}$ memristive technology non-idealities.\\ \newline
\newline
The tests performed to gather the data in the non-ideality database used during training for tuning imprecision, biasing scheme effects and finally HRS or LRS stuck devices are described in the following subsections. In the case of the weight disturbances induced by the biasing scheme effect, fitting normal distributions to the data was unfortunately impossible. The Shapiro-Wilk tests for normality of the distributions fitted to the data usually returned an alpha value of less than 0.05, causing the null hypothesis to be rejected \cite{SHAPIRO1965}. A sample from this distribution does not approximate a recorded disturbance instance well. The approach taken instead was to sample a recorded conductance change from a portion of the database consisting of only raw data. This is different from sampling from a fitted distribution, which is done for the conductance tuning imprecision. In all of the tests, the conductance tuning procedure followed the pulse tuning technique presented in \cite{Alibart2012} with a 1\% tuning tolerance. \newline
\newline
\textbf{Device conductance tuning imprecision} 
\newline

As shown in Figure \ref{fig:figure 1}a, the device-level conductance tuning imprecision is considered in the training process. The studies undertaken to characterize it were realized by dividing the achievable resistive ranges of all 29 studied devices into eight different equidistant levels and repeatedly performing a tuning/untuning procedure. Figure \ref{fig:conductance_tuning} illustrates how the statistical database was established, along with an example. Figure \ref{fig:conductance_tuning}a shows the experimental data behind one instance of the procedure for a target conductance of 125 $\mu S$. The distribution of read values after the conductance reached and stabilized around the target conductance is shown in Figure \ref{fig:conductance_tuning}b.  This tuning/untuning procedure was repeated 40 times per level and per device. In Figure \ref{fig:conductance_tuning}c and d, the  data points corresponding to the data distribution in Figure \ref{fig:conductance_tuning}b are depicted in red and are evaluated as follows: \newline
\begin{equation}
\label{eq:fig2}
\begin{split}
&(x_{c}, y_{c}) = \left(g_{target}, \sigma_{program}\right) = \left(125\mu S, 0.57\% \right),\\
&(x_{d}, y_{d})=\left(g_{target}, \frac{\mu_{program} - g_{target}}{g_{target}} * 100\right) \\ 
&= \left(125\ \mu S, -0.424\%\right).
\end{split}
\end{equation} \newline
 \begin{figure*}[]  
    \centering
    \includegraphics[width=1.0\linewidth]{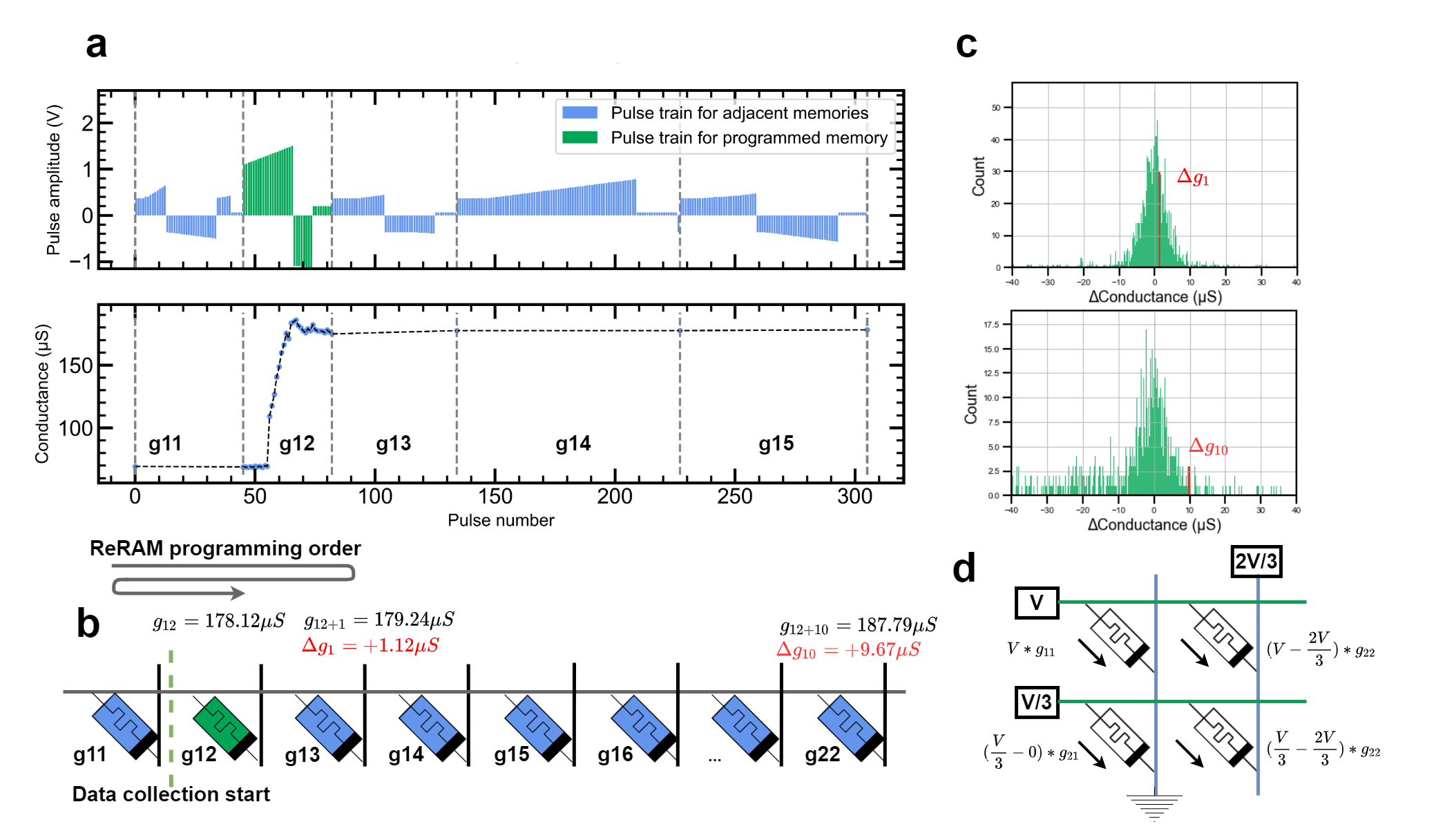}
    \caption{\textbf{Depiction of the procedure for gathering data on device conductance changes caused by the biasing scheme.} \textbf{a.}\,The top graph depicts the pulse train seen by a device during its own programming (in green) and during the programming of its neighbors (in blue). The bottom graph shows the conductance evolution of a device following the pulse trains displayed in the top graph. The device changes the conductance significantly during its programming cycle and changes it slightly during its neighbors' programming cycle. \textbf{b.}\, Color-coded depiction of a section of the crossbar studied in panel \textbf{a} with numerical values for the measured conductance change seen by the device after the programming cycle of other memories. \textbf{c.} Histogram of the recorded conductance changes with the bins containing the changes illustrated in panel \textbf{b} highlighted in red. The top graph contains the conductance changes when one additional device is programmed and the bottom graph contains the conductance changes when ten additional devices are programmed. \textbf{d.} Depiction of the V/3 and 2V/3 biasing schemes for programming devices.}
    \label{fig:bias_scheme_effect}
\end{figure*}  
 Figure \ref{fig:conductance_tuning}c shows the standard deviation of the programing precision as a function of the conductance state of the devices tested, where the values are expressed in terms of a percentage of the desired conductance value, while Figure \ref{fig:conductance_tuning}d shows the offset between the mean of the read conductance values after the tuning of the devices and their target conductance values. Since the devices are always tuned from a low conductive level to a high conductive level, the mean offset of the achieved conductance from the ideal target conductance is always negative. Therefore, the transfer on a crossbar should be performed after all devices are RESET in order to match this replicable offset, meaning that all devices should start from the off state. \newline
 
 During training, when the network weights are transformed into software conductances \textit{$g_{s+}$} and \textit{$g_{s-}$} for signed weights or \textit{$g_{s}$} for unsigned weights, a weight is sampled from a normal distribution centered around these values with a standard deviation that depends on the conductive level of the transformed weight. Figure \ref{fig:conductance_tuning}c shows the function $f(g)$ defining this standard deviation relationship. This allows the network to take into account the correlation that is often observed between the conductive level and variability \cite{Dalgaty2021}. An offset is then sampled from a distribution $N\left(\mu_{off}, \sigma_{off}\right)$, which is depicted in Figure \ref{fig:conductance_tuning}d, and added to the sampled conductance value. 
 Because the test is performed by reading ten times before considering a memory to be programmed, the read variability on devices is also taken into account. An experiment was undertaken to isolate the read variability of the characterization equipment with respect to the variability of the measured values; it was found that the reading variability stemming from the instrument is one order of magnitude lower than that of the experimental data (Supplementary Figure \ref{fig:b1500_varia}.) \newline
\begin{figure*}[t]  
    \centering
    \includegraphics[width=1.0\linewidth]{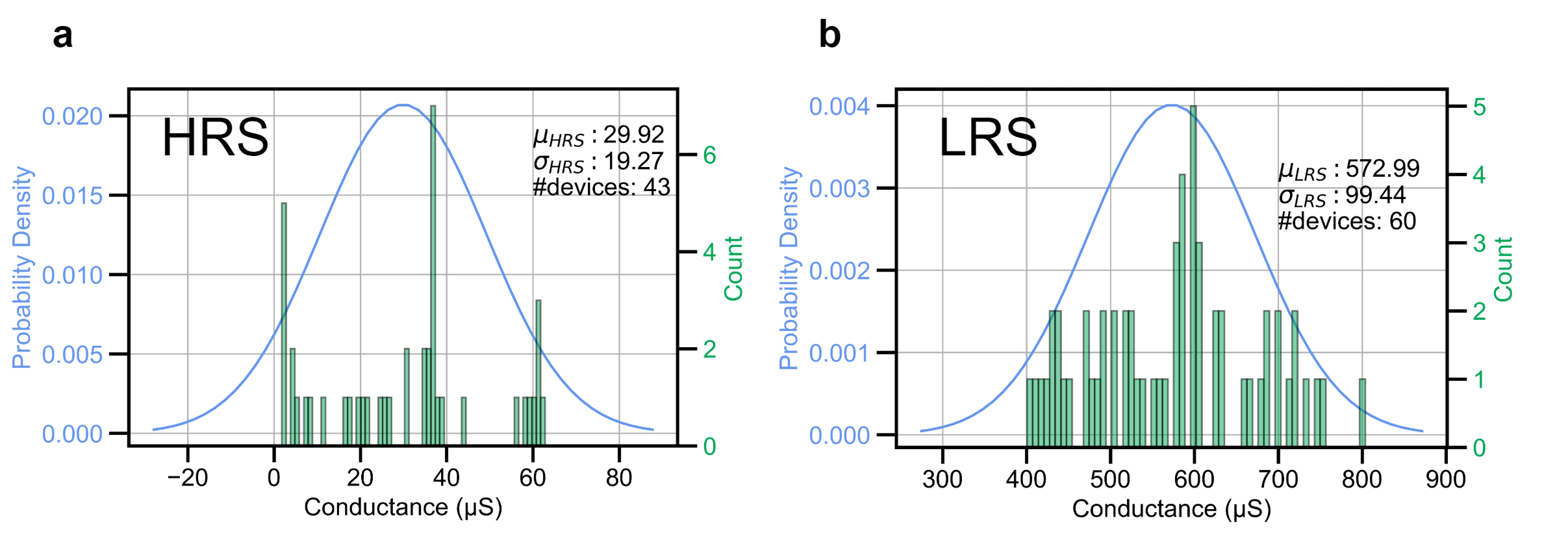}
    \caption{\textbf{Distribution of faulty ReRAM.} \textbf{a.} Recorded conductance levels of dead crosspoints stuck in an HRS. The Shapiro-Wilk p-value is ~0.006 and therefore the hypothesis of the normality of the distribution is rejected. \textbf{b.} Recorded conductance levels of dead crosspoints stuck in an LRS. A device is considered stuck when it fails to change its conductance state after the application of 75 successive pulses of maximum SET or RESET amplitudes (2.0 V or -2.4 V). }
    \label{fig:LRS_stuck}
\end{figure*}  
\textbf{Conductance disturbances caused by biasing scheme effect}\newline

Programming devices on a passive crossbar require the use of a biasing scheme to reduce sneak path currents. The V/3 scheme was used in this study since all devices other than the addressed memory will see the same voltage of V/3. This standardizes and simplifies the process of adding variability to the weights. It also makes it possible to use higher-amplitude voltages to tune the devices without affecting neighboring devices significantly. In order to measure the change in the conductance of a device due to the programming of its neighbors, we tuned devices from three different crossbars, for a total of 144 devices (64 + 64 + 16), to random values within the expected conductance ranges $G_{R}$. The disturbance of a device, after the tuning cycles for devices that are tuned after itself are complete, is recorded in the database. This disturbance is due to the V/3 biasing scheme. A thorough example is displayed in Figure
\ref{fig:bias_scheme_effect}a and b. The conductance changes caused by the biasing scheme accumulate through the crossbar programming, which can be observed by comparing the disturbance due to the programming of the neighbors of the device that is tuned first with that of the devices that are tuned last. As such, the database of recorded disturbance measurements is separated into sub-databases for each number of subsequent devices tuned after the programming of the current device. Figure \ref{fig:bias_scheme_effect}c shows two such sub-databases, one for the scenario in which one additional device is tuned and one for the scenario in which ten additional devices are tuned. The device that is tuned last is the bottom-rightmost weight of our neural network matrix when it is transferred from top to bottom and left to right. It will see no conductance change due to the biasing scheme since it is the last-tuned device. The devices stuck in an HRS or LRS or that could not reach their target conductances were also not added in this database. If they had been included, the database would have been polluted by stuck or ineffective devices, which would have biased it towards no conductance change. \newline
As can be seen in Figure \ref{fig:bias_scheme_effect}c, a tail of recorded crosstalk effects is more prevalent for negative values. This is an indirect consequence of the asymmetrical device switching mechanism between the SET and RESET pulses. More negative pulses are required to reset devices than positive pulses. Thus, devices will, on average, see more successive negative pulses than positive pulses. This phenomenon is worsened by the fact that devices are individually  affected differently by negative pulses, with more sensitive devices leading to greater conductance changes.  A conductance change of up to 60 $\mu$S was allowed in the sub-databases; this change is six times larger than the lowest possible conductance weight. The cases in which the disturbance is greater than this value usually correspond to an HRS or LRS failure. 

During the training of the neural network, a random offset sampled from these sub-databases correlated to the positions of the weights in the crossbar is added to each weight. The use of the V/3 scheme, as shown in Figure \ref{fig:bias_scheme_effect}d, makes it possible to proceed in this way. This ensures that the first weights in the matrices will see more V/3 amplitude pulses and will, on average, have a greater variability than the last devices that are programmed.\newline
\newline
\textbf{HRS-LRS stuck device substitution}
\newline
\newline
Some devices are stuck in a low or high resistive state due to defects. Regardless of the DC voltage or the pulse amplitudes that are applied to such devices, their conductive levels will never be tunable. As shown in Supplementary Figure \ref{fig:retention_test}, LRS failures include retention failures. Such failures are caused by temporal effects and cause the device conductance to drift over time to a typically higher conductance state. HRS failures may occur for devices that were not forming free or devices with a very narrow achievable conductance range lying outside the desired range. To account for these device faults, percentages of devices \textit{x} and \textit{y} corresponding to HRS and LRS stuck devices, respectively, have both positive ($g_{s+}$) and negative ($g_{s-}$) software weight components that are replaced by a randomly sampled failed device conductance. The percentages \textit{x} and \textit{y} are fixed throughout the training, but the affected weights are randomized for every batch of input data. This implies that certain weights will be doubly affected by this substitution if their positive and negative components are both assigned as failures. \newline
The stuck devices' conductances are sampled from two distributions based on the characterization of LRS and HRS stuck devices made on five  8$\times$8 crossbars (320 devices). The conductance range of devices stuck in an HRS is much smaller than that of devices stuck in an LRS; HRS devices are defined as devices stuck below our HRS resistance of 100$\mu$S. The distribution of failed HRS state devices can be seen in Figure \ref{fig:LRS_stuck}a, while the distribution of failed LRS devices can be seen in Figure \ref{fig:LRS_stuck}b. Since fitting the normal distribution is inappropriate for the HRS distribution, as it fails the Shapiro-Wilk normality test, HRS stuck device faults are uniformly sampled from the range [100 $\mu$S, 10 $\mu$S]. The impact of devices stuck below 10 $\mu$S will be more negligible than the impact of those stuck at <100 $\mu$S and >10 $\mu$S. Indeed, devices stuck in a higher conductance state will have a greater impact on the summation of the currents during the analog matrix multiplication. Gradient descent is deactivated for the weights affected by HRS or LRS stuck weights. Typically, the fabrication yields of working devices on crossbars for state-of-the-art memristive devices are above 95\% \cite{Bayat2018_95}, with some articles reporting near perfect crossbars with yields of 99\% \cite{Kim2021} or even 100\% \cite{100yield}. The percentages of HRS and LRS stuck devices should be chosen to reflect this fact.  In fact, in general, the values of these parameters should be chosen as conservatively as possible so that the network converges to a solution that is robust to as many device failures as possible. Of course, replacing too many weights with random values at the extremities of the conductance range and therefore at the extremities of the overall weight range will prevent the network from converging to a solution, making the training impossible. This makes these percentage values tunable hyperparameters. 
 \newline
\newline
\textbf{Mathematical expression}
\newline
\newline
Overall, the mathematical equation representing the sampling of each of the devices for either positive or negative weight components is equation \ref{eq:1}:

\begin{equation}
\label{eq:1}
\begin{split}
\scriptstyle
&N\left(\mu_{w\pm},{\sigma_{w\pm}}^2\right)\\&=N\left(g_{s\pm},f(g_{s\pm})^2\right)+N\left(\mu_{off}, \sigma_{off}^2\right)+ g_{adj}(n_{d}),\\
&=N\left(g_{s\pm}+\mu_{off}+g_{adj}(n_{d}), \sigma_{off}^2+f(g_{s\pm})^2\right),
\end{split} 
\end{equation}\newline
where $g_{s}$ is the direct conductance equivalent of the weight after a linear transformation between the two ranges (see Supplementary Note 1). The weight's range is reevaluated for every batch during training since the maximum and minimum weight values are subject to change as the training progresses. $ N\left(\mu_{o f f}, \sigma_{off}\right)$ is the sampled conductance tuning imprecision offset and $g_{adj}(n_{d})$ is the sampled biasing scheme effect from programming adjacent devices with respect to the number of additional devices $n_{d}$ programmed after the current device. The variability associated with a device in training is dependent on both its target conductance level for both negative and positive weight components and the position of the devices in their respective crossbar. Every weight matrix is presumed to be programmed on a tiled crossbar or on a sub-section of a crossbar.

The network learns to expect more variability from the first devices that are programmed because of the biasing scheme effects and from devices at low conductance values and will adjust the importance of each weight accordingly. The substitution of random failed weights acts as a form of aggressive regularization, improving the network's generalization. As can be expected, training a neural network with stochastic jumps in weight values leads to a longer training time compared to a regular neural network, as is the case for Bayesian neural networks. However, this constant sampling also enables the network to be more robust to overfitting \cite{watanabe_2009}.

\textbf{Comparison with regular neural network}
\begin{figure}[h] 
    \centering
    \includegraphics[width=1.0\linewidth]{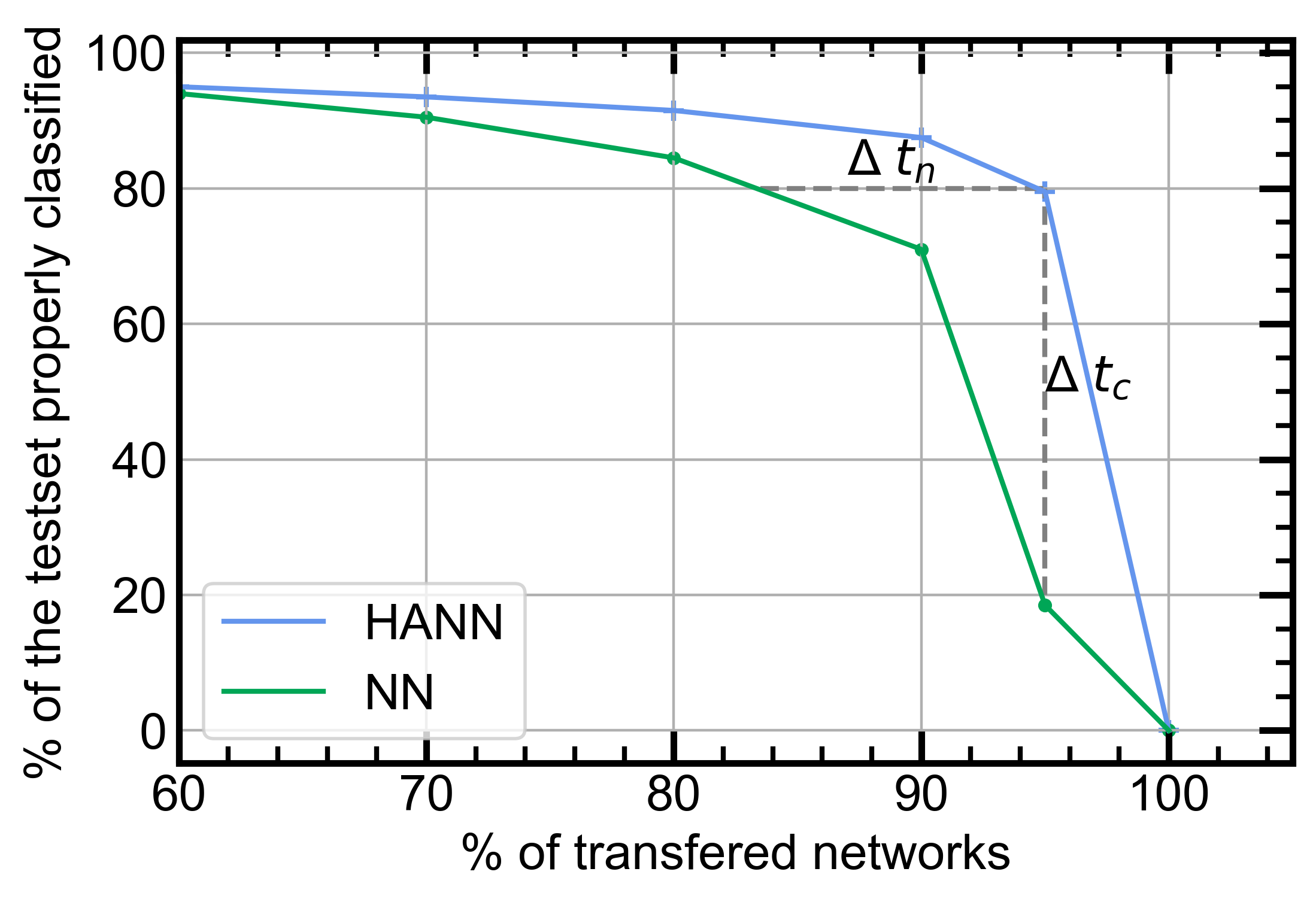}
    \caption{\textbf{Comparison of the minimal accuracy over 10,000 simulated transfers achievable for the test set.} None of the datapoints are successfully classified by 100\% of the simulated neural networks due to the often aggressive substitution of weights by failing devices, large crosstalk effects and imprecise tuning. We can see that more of the test set is classified with a higher accuracy for the HANN (depicted in blue) compared to the NN (in green). For instance, 80\% of the test set is properly classified by 83.5\% of the simulated transferred regular networks vs ~95\% of the simulated transferred hardware-aware networks. This corresponds to an increase of $\Delta t_{n}$ = 11.5\%. Reading the graph in the opposite way, we see that 95\% or more of the simulated regular networks are capable of properly classifying only 18.5\% of the test set, while 95\% of the hardware-aware networks are capable of properly classifying 79.5\% of the test set. This corresponds to an increase of $\Delta t_{c}$ = 61\%. }
    \label{fig:compNN}
\end{figure}  

\begin{figure*}[t!]  
    \centering
    \includegraphics[width=1.0\linewidth]{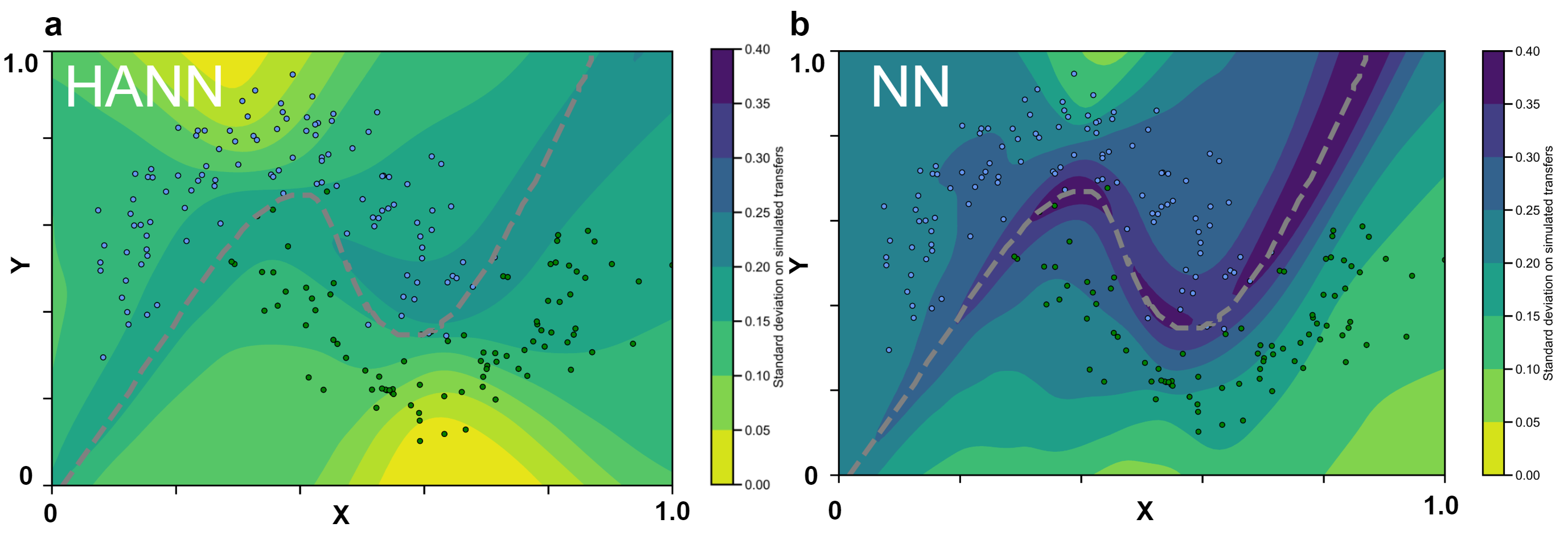}
    \caption{\textbf{Heatmap of the variability of classifications with respect to the coordinates (x, y) used as input across the data space.} \textbf{a.} Variability of the hardware-aware neural network. \textbf{b.} Variability of the regular neural network.  The inference is repeated 1,000 times and the mean and standard deviation of the classified results (always 0 or 1) are obtained. A mean smaller than or equal to 0.5 corresponds to the blue class and a mean greater than 0.5 corresponds to the green class. The standard deviation is depicted in the heatmap, with a deeper blue corresponding to more variability across classifications. Note that both subfigures share the same color scale to make it possible to visually compare them. Most data points for the regular neural network lie within regions of greater variability than for the HANN. A dashed gray line that separates the two classes with a precision of 98\% is provided for reference.}
    \label{fig:results}
\end{figure*}  

A simple half-moon problem was considered to compare the performance of our hardware-aware training approach with that of a regular neural network trained without taking into account hardware non-idealities. The two networks have the same learning rate $\alpha$ of 0.01, and they use an Adam optimizer \cite{ADAM} with a momentum of 0.9. The batch size used was 256. A neural network with one hidden layer with eight neurons is considered, and the sigmoid activation function is used (as depicted in Figure \ref{fig:figure 1}b).  The cost function used is the binary cross-entropy loss \cite{Aldrich1997}. During training, hyperparameter values of 0.5\% were chosen for both the LRS-stuck and HRS-stuck substitutions. For the first layer, a recorded average of 3.1\% of neural network weights were affected by random substitutions of LRS and HRS stuck devices during training. A half-moon classification problem training set of 875 data points and a test set of 200 data points were randomly generated. The neural network (NN) is trained without any consideration of the variability and then evaluated with the stochastic effects added to compare its performance to that of the hardware-aware NN (HANN). This is repeated \textit{n} times to determine what percentage of the test set could be classified well over a given percentage of the \textit{n} simulated transferred networks. These results are expressed in Table \ref{tab:results} in Supplementary Note 4. A comparison of the accuracies is also displayed in Figure \ref{fig:compNN}. Interpreting these results, we see that 79.5\% of the test set can be properly classified by at least 95\% of the simulated transferred hardware-aware networks. On the other hand, only 18.5\% of the test set can be properly classified by at least 95\% of the simulated transferred regular neural networks. We observe that 87.5\% of the test set can be properly classified by at least 90\% of the simulated transferred hardware-aware networks, while 71\% of the test set can be properly classified by at least 90\% of the regular networks. The hardware-aware neural network training techniques contribute significantly to maintaining a high consistency and accuracy between simulated transferred networks. 
Figure \ref{fig:results} offers a graphical interpretation of the difference in the relative variability by showing the results of the test set classifications over 1,000 simulated transfers for each of the two networks, along with a heatmap representation of the variability with respect to the entire data space for the half moons dataset. Note that this is for a single instance of a hardware-aware-trained NN and a regularly trained NN. The variability, depicted by a deeper blue, is greater for the regular neural network and, as expected, is also greater at the frontier between the two classes.

\section{Conclusion}

 It was demonstrated in this paper that the use of hardware-aware training techniques can make neural networks significantly more robust to their often difficult transfer onto passive crossbar arrays. This protocol integrates experimentally measured non-idealities such as the conductance tuning imprecision, biasing scheme effects and device failures into the neural network training. \newline
 
 A classification task using the half moons dataset reveals that our hardware-aware training protocol results in a 95\% accuracy across simulated transfers for 79.5\% of the test set, compared to only 18.5\% for the regular neural network. The hardware-aware training procedure could enable the use of passive crossbars despite their non-idealities for inference tasks and proposes a parallel solution to the usual fabrication process improvements. It may serve as a middle ground between fabrication and software solutions to accelerate practical, real-life applications of memristive devices. Furthermore, setting up a custom hardware-aware neural network architecture using preexisting libraries proves to be relatively straightforward and inexpensive in terms of hardware.\newline
 
 While the devices studied here are TiO$_{\textrm{2}}$ passive crossbars, techniques like this are applicable to different memristive technologies as long as a new database is built using the same procedures presented in this article. A future study should focus on evaluating the scalability of this approach, with respect to the advantages (robustness, accuracy) and disadvantages (training time, variability-induced local minima). The potential of using this technique for more complex datasets should also be evaluated. Faster convergence during training could most likely be achieved by integrating the variability into the gradient descent procedure. This model could be improved by adding additional variability sources and by developing versatile libraries that make hardware non-idealities compatible with neural network requirements. The philosophy behind this work is to realistically render compatible in-memory computing for neural networks with the current non-negligible hardware non-idealities present in passive crossbar arrays. This demonstration, based on experimental data, is a step forward in making the use of passive crossbars for neural network matrix-vector multiplications viable in spite of their inherent shortcomings.

 \section*{Methods}\label{Methods}
\textbf{Tuning algorithm and characterization details} \newline
The pulse tuning algorithm presented in \cite{Alibart2012} was used. A memory was considered programmed when ten successive reading pulses returned a value within a tolerance around the target. The writing pulses were incremented from [0.8 V, 2 V] for SET pulses and from [-1.1 V, -2.4 V] for RESET pulses and had a duration of 200 ns. The read pulses had a duration of 10 $\mu$s and an amplitude of 0.2 V, while the current measurement range was 100 $\mu$A. The devices considered were forming free. Figure \ref{fig:conductance_tuning}a depicts the conductance evolution using the pulse conductance tuning procedure. Devices that failed before the completion of the testing procedure were kept in the database, as this corresponds to a realistic scenario of a device's behavior before imminent failure, while devices that failed before testing began were kept in the HRS-stuck or LRS-stuck databases. More information regarding the characterization setup can be found in Supplementary Figure \ref{fig:supp_elec_setup} and Supplementary Note 2. A graph of hysteresis curves under DC stimulation for 129 of our devices is also available in Supplementary Figure \ref{fig:hysteresis}. \newline
\newline
\textbf{Fabrication of CMOS-compatible TiO${_\textrm{2}}$-based memristor crossbar} \newline
The samples used for measurements were prepared as described in a previous publication \cite{mesoudy2021CMOS}. First, a 600-nm-thick SiN layer was deposited on a Si substrate using low-pressure chemical vapor deposition (LPCVD). Electron-beam lithography (EBL) was used to pattern ~400-nm-wide bottom electrodes, which were etched 150 nm deep using an inductively coupled plasma etching process with CF$_4$/He/H$_2$ chemistry. The trenches were filled with 600-nm-thick TiN, deposited by reactive sputtering, and polished using chemical-mechanical polishing (CMP) to remove the excess TiN and planarize the samples. This resulted in embedded TiN electrodes within the SiN layer. The active switching layer was composed of 1.5 nm of Al$_2$O$_3$ and 15 nm of TiO$_2$, which were deposited using atomic layer deposition and reactive sputtering, respectively. EBL and plasma etching with BCl$_3$/Cl$_2$/Ar chemistry were used to deposit and pattern ~400-nm-wide top electrode lines with Ti (10 nm)/TiN (30 nm)/Al (200 nm). The switching layer outside the crossbar region was etched using the same process to suppress line-to-line leakages and open the ends of bottom electrodes. The crossbars were then encapsulated within a 500-nm-thick PECVD SiO$_2$ layer, and the contacts on the open ends of the top and bottom electrodes were etched using plasma etching with CF$_4$/He/H$_2$ chemistry. Ti (10 nm)/Al (400 nm) contact pads for wiring were deposited and patterned using EBL and plasma etching with BCl$_3$/Cl$_2$/Ar chemistry. Finally, the sample was subjected to a rapid thermal annealing process at 350$^\circ$C in N$_2$ gas for an unspecified duration. \newline
\noindent\textbf{Data availability}
 The hardware-aware training library, along with some of the experimental data, can be found in a GitHub repository that has been made publicly available at \cite{DroletHARDWAREAWARE}. The rest of the data that support this work are available from the corresponding author upon reasonable request. 
\linebreak
\linebreak

\textbf{Acknowledgements}
This work was supported by the Natural Sciences and Engineering Research Council of Canada (NSERC) and by the Fonds de Recherche du Québec Nature et Technologie (FRQNT). We acknowledge financial support from the EU: ERC-2017-COG project IONOS (\# GA 773228). We would like to acknowledge Jonathan Vermette, Vincent-Philippe Rhéaume and the LCSM and LNN cleanroom staff for their assistance with the electrical characterization setup and nanofabrication process development. LN2 is a joint International Research Laboratory (IRL 3463) funded and co-operated in Canada by Université de Sherbrooke (UdeS) and in France by CNRS as well as ECL, INSA Lyon, and Université Grenoble Alpes (UGA). It is also supported by the Fonds de Recherche du Québec Nature et Technologie (FRQNT). We would also like to thank the IEMN cleanroom engineers for their help with the device fabrication.
\\

\noindent\textbf{Author contributions}
P.D. performed most of the experiments, designed the neural networks, analyzed and post-processed the measurements and created the characterization setup. V.Y. helped in designing the neural network architecture. P.A.M. contributed to the experimental setup and along with M.V. helped with the characterization. R.D., J.A.Z, P.G., S.E. and M.V. fabricated the memristor devices. Y.B., F.A., S.E., S.W. and D.D supervised the project. P.D. wrote the manuscript with input from all of the authors.\\
\linebreak
\textbf{Competing interests}
The authors declare no competing interests.
\bibliographystyle{IEEEtran}
\bibliography{references}
\pagebreak
\clearpage
\section*{Supplementary Materials}\label{Methods}
\textbf{Supplementary Note 1: Mathematical expression of range conversions between weights and conductances}
\newline 
The following equations represent the conversion of the weights to the conductance range, the addition of the noise from the reparametrization trick and the conversion back to the initial weight range. In these equations, $\phi_{\min }=\min (\phi)$ is the minimal weight value,  $\phi_{\max }=\max(\phi)$ is the maximal weight value and $\phi_{\text {absmax }}=\max (|\phi|)$ is the maximal absolute value of the weights. $g_{max} = 400 \mu$S and $g_{min} = 100 \mu$S, as described previously in this article. As shown in equation \ref{eq:W-DIV-pos-neg}, the weights are separated into negative and positive components, with all weight components of the opposite sign being set to 0 and negative weight values being converted into positive values. The weights are then converted through min-max scaling to the conductance range, as shown in equation \ref{eq:conv_to_cond}. An epsilon parameter that depends on the obtained positive and negative component conductance values is added to the weights, which are then converted back into the initial range, as shown in equation \ref{eq:final_conv}. The value of the final $\epsilon$ parameter for the reparametrization trick, as shown in Figure \ref{fig:figure 1}b, is therefore represented by equation \ref{eq:final-eq-epsilon}: 
\begin{equation}
\label{eq:W-DIV-pos-neg}
\begin{multlined}
\phi \mapsto \phi_{+}-\phi_{-},\\
\phi_{+}=\phi \geq 0,\\
\phi_{-}=-\phi \geq 0,\\
\end{multlined}
\end{equation}
\begin{equation}
\label{eq:conv_to_cond}
\phi_{\pm} \mapsto \frac{\phi_{\pm}-0}{\phi_{\text {absmax }}-0} \times\left(g_{\max }-g_{\min }\right)+g_{\min }=\phi_{g \pm}, 
\end{equation}
\begin{equation}
\phi_{g \pm}=\phi_{g \pm}+\varepsilon_{g\pm}\left(\phi_g{\pm}\right),
\end{equation}
\begin{equation}
\label{eq:final_conv}
\begin{multlined}
\phi_{g+}-\phi_{g-}= \Delta \phi_g, \\ 
\Delta \phi_g \mapsto \frac{\left(\phi_{g+}-\phi_{g-}\right)-\left(g_{\min }-g_{\max }\right)}
{\left(g_{\max }-g_{\min }\right)-\left(g_{\min }-g_{\max }\right)} \\
\times\left(\phi_{\max }-\phi_{\min }\right)+\phi_{\min }=\phi^{\prime},
\end{multlined}
\end{equation}
\begin{equation}
\label{eq:final-eq-epsilon}
\epsilon(\phi) = \phi - \phi'.
\end{equation}

\setlength{\parindent}{0em}
\textbf{Supplementary Note 2: Electrical characterizations} The electrical characterizations performed for the conductance tuning experiments were carried out with a Keysight B1500 parameter analyzer. The waveform generator/fast measurement unit (WGFMU) module was used, along with a custom C\texttt{++} script based on a Keysight library \cite{MounyB15002021}. The instruments were controlled through a GPIB-USB connection. The crossbar programming and characterization were performed using a Keithley S4200 system attached to a custom pulse divider PCB and connected to a Keithley 707B switch matrix. The signals were then multiplexed onto a wire-bonded chip resting on another printed circuit board (PCB) platform, as depicted in Supplementary Figure \ref{fig:supp_elec_setup}. A custom C library was created to be used with the \textit{KULT}, \textit{KITE} and \textit{KCON} software suite on the Keithley. \newline
\textbf{Supplementary Note 3: Comparative table of the results of the neural networks}
\begin{table}[H]
\centering
\caption{Relative accuracies of every data point in the test set for both the hardware-aware neural network (HANN) and the regular neural network (NN) over 10000 simulated transfers. These points correspond to the points in Figure \ref{fig:compNN}.}
\label{tab:results}
\begin{tabular}{|l|l|l|}
\hline
\hline
\textbf{\begin{tabular}[c]{@{}l@{}}Percentage of\\ simulated networks\\that correctly classify\\
the datapoints(\%)\end{tabular}} & \multicolumn{1}{c|}{\textbf{\begin{tabular}[c]{@{}c@{}}NN \\ \#data points\end{tabular}}} & \multicolumn{1}{c|}{\textbf{\begin{tabular}[c]{@{}c@{}}HANN\\ \#data points\end{tabular}}} \\ \hline
100                                                                              & 0                                                                                         & 0                                                                                         \\ \hline
95 $\leq$ x \textless 100                                                     & 37 (18.5\%)                                                                               & 159 (79.5\%)                                                                              \\ \hline
90 $\leq$ x \textless 95                                                      & 105 (52.5\%)                                                                              & 16 (8\%)                                                                                  \\ \hline
80 $\leq$ x \textless 90                                                      & 27 (13.5\%)                                                                               & 8 (4\%)                                                                                   \\ \hline
70 $\leq$ x \textless 80                                                      & 12 (6\%)                                                                                  & 4 (2\%)                                                                                   \\ \hline
60 $\leq$ x \textless 70                                                      & 7 (3.5\%)                                                                                 & 3 (1.5\%)                                                                                 \\ \hline
50 $\leq$ x \textless 60                                                      & 6 (3\%)                                                                                   & 2 (1\%)                                                                                   \\ \hline
x $\leq$ 50                                                                & 6 (3\%)                                                                                   & 8 (4\%)                                                                                   \\ \hline
\textbf{Total}                                                                   & 200 (100\%)                                                                               & 200 (100\%)                                                                               \\ \hline

\end{tabular}
\end{table}
\newpage
\begin{figure*}[]  
    \centering
    \includegraphics[scale = 0.35]{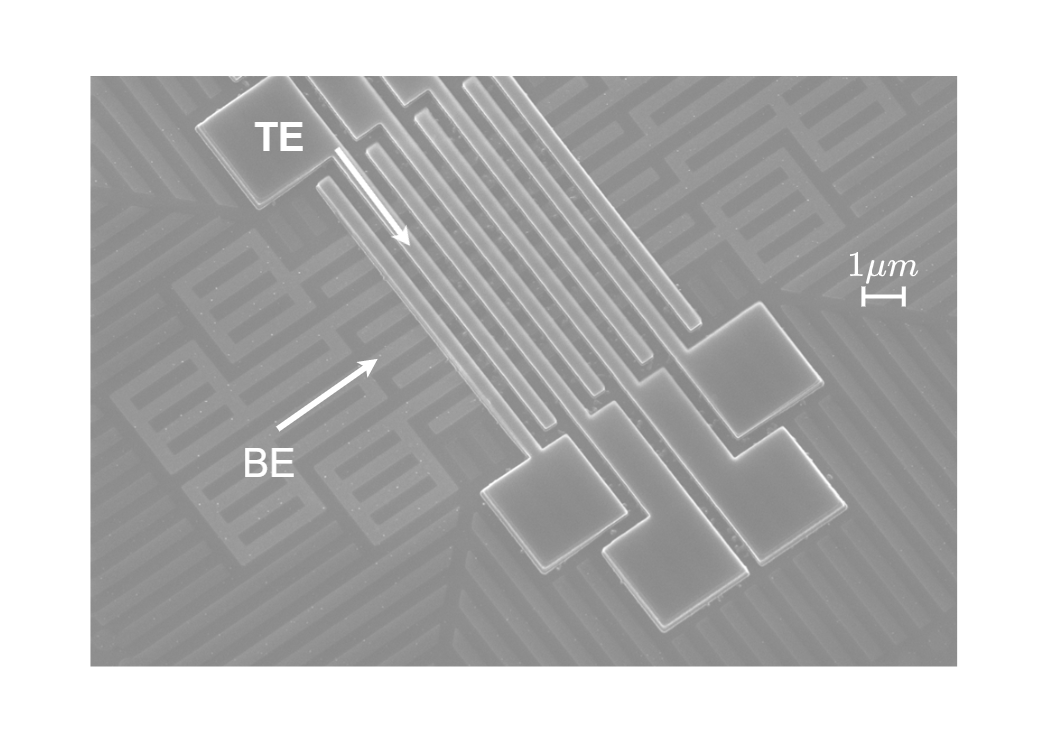}
    \caption{\textbf{Picture of a crossbar after top electrode patterning.} Close-up view of the crossbar structure of our TiO${_\textrm{2}}$-based devices after top electrode fabrication. The top electrode (TE) and bottom electrode (BE) are indicated on the picture. The pattern on the bottom electrode is implemented in order to improve the chemical mechanical polishing (CMP) employed during the damascene process.}
    \label{fig:supp_sem_closeup}
\end{figure*}  
\begin{figure*}[t]  
    \centering
    \includegraphics[scale = 0.5]{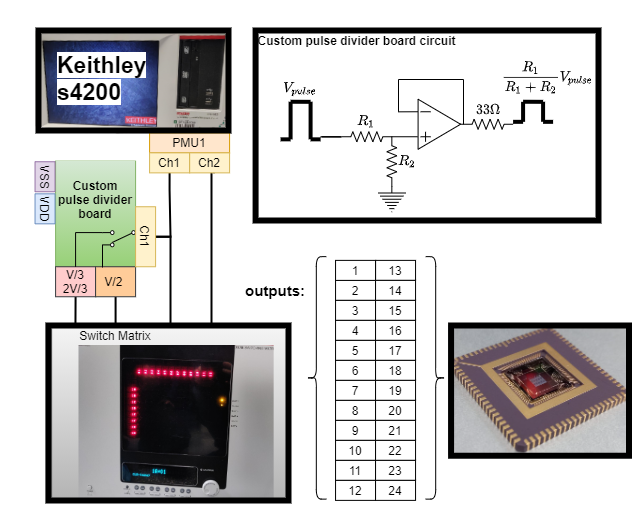}
    \caption{\textbf{Electrical characterization setup for crossbar measurements.} A simple current follower circuit is used to divide the pulse into fractions of itself (which can be V/2 or V/3 and 2V/3). The outputs of the switch matrix are then connected to a wire-bonded chip through a series of coaxial cables.}
    \label{fig:supp_elec_setup}
\end{figure*}  
\begin{figure*}[t]  
    \centering
    \includegraphics[scale = 0.7]{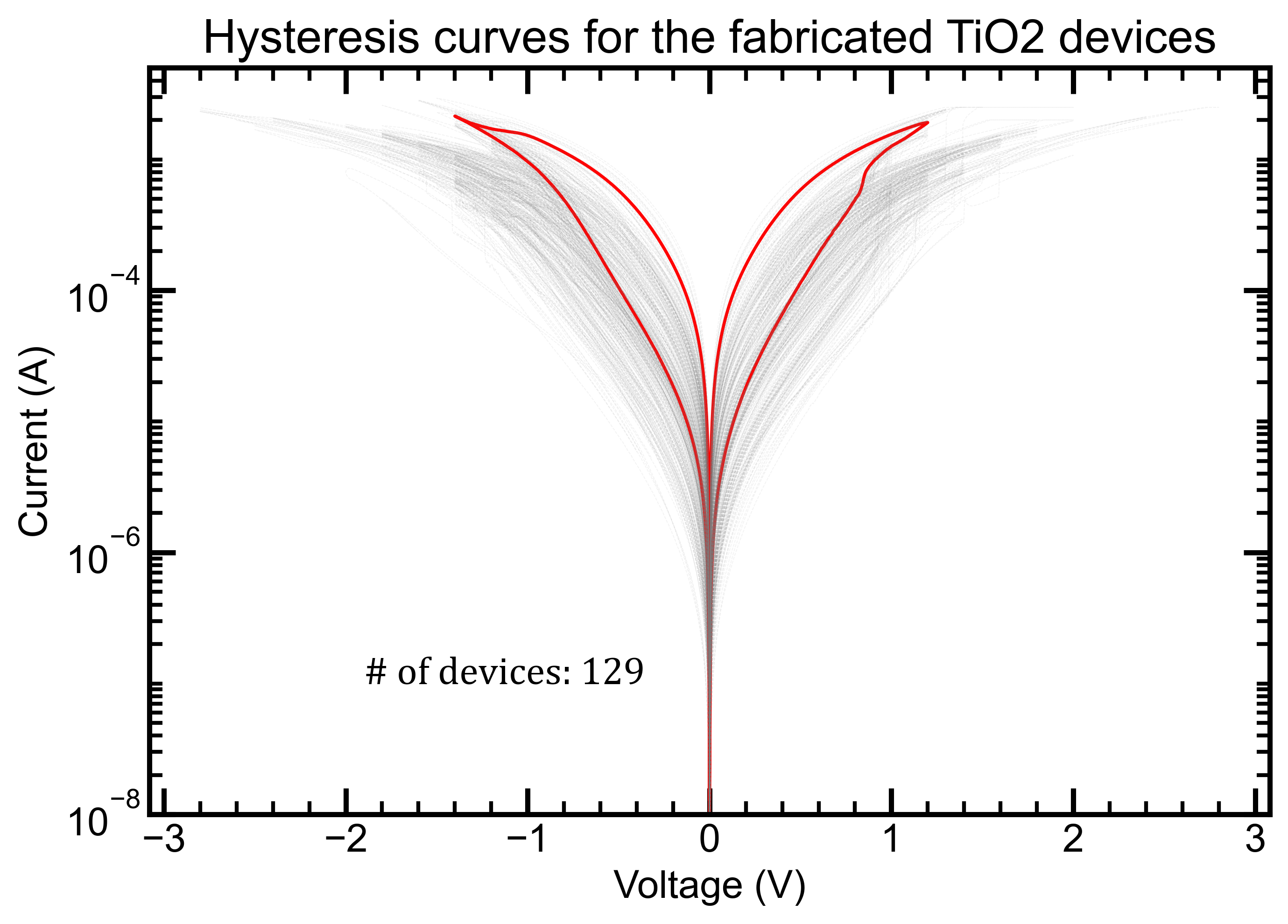}
    \caption{\textbf{Representation of the DC cycling of 129 of our devices (in crosspoint configuration).} Devices were cycled using DC voltages reaching up to -3 V for RESET and 3 V for SET. One device's cycling is highlighted in red for reference. Only working devices are considered. }
    \label{fig:hysteresis}
\end{figure*}  
\begin{figure*}[t]  
    \centering
    \includegraphics[scale = 0.40]{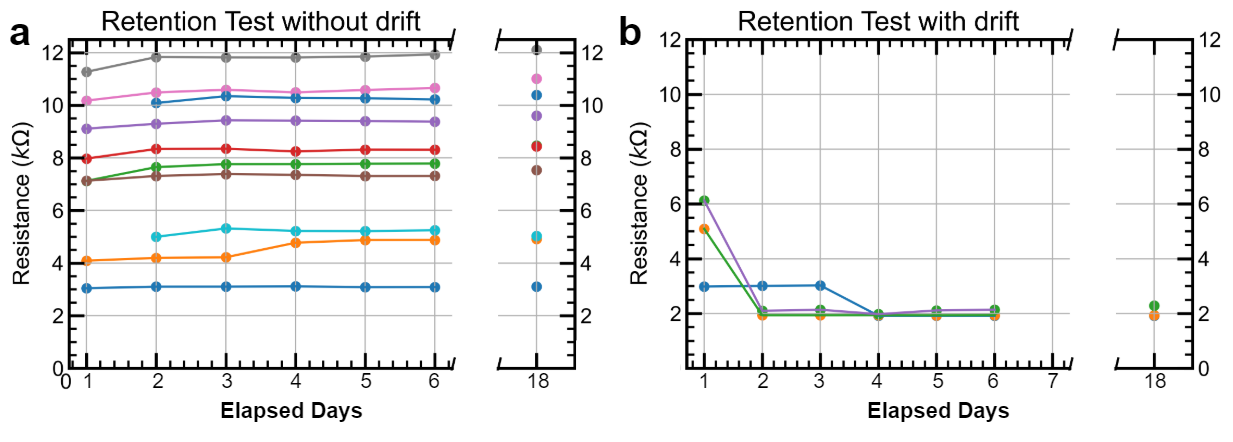}
    \caption{\textbf{Endurance test drift leading to LRS failure.} \textbf{a.}\, Conductance evolution through time of devices that did not drift for the duration of the test \textbf{b.}\, Conductance evolution through time of devices that drifted during the test.  Devices were programmed to different values within their resistive ranges and their resistances were read once per day for 18 days; three of the devices showed drift towards the LRS. A read voltage of 0.2 V was used. This demonstrates that retention failures are LRS-type failures.}
    \label{fig:retention_test}
\end{figure*} 

\begin{figure*}[t]
    \centering
    \includegraphics[scale=0.7]{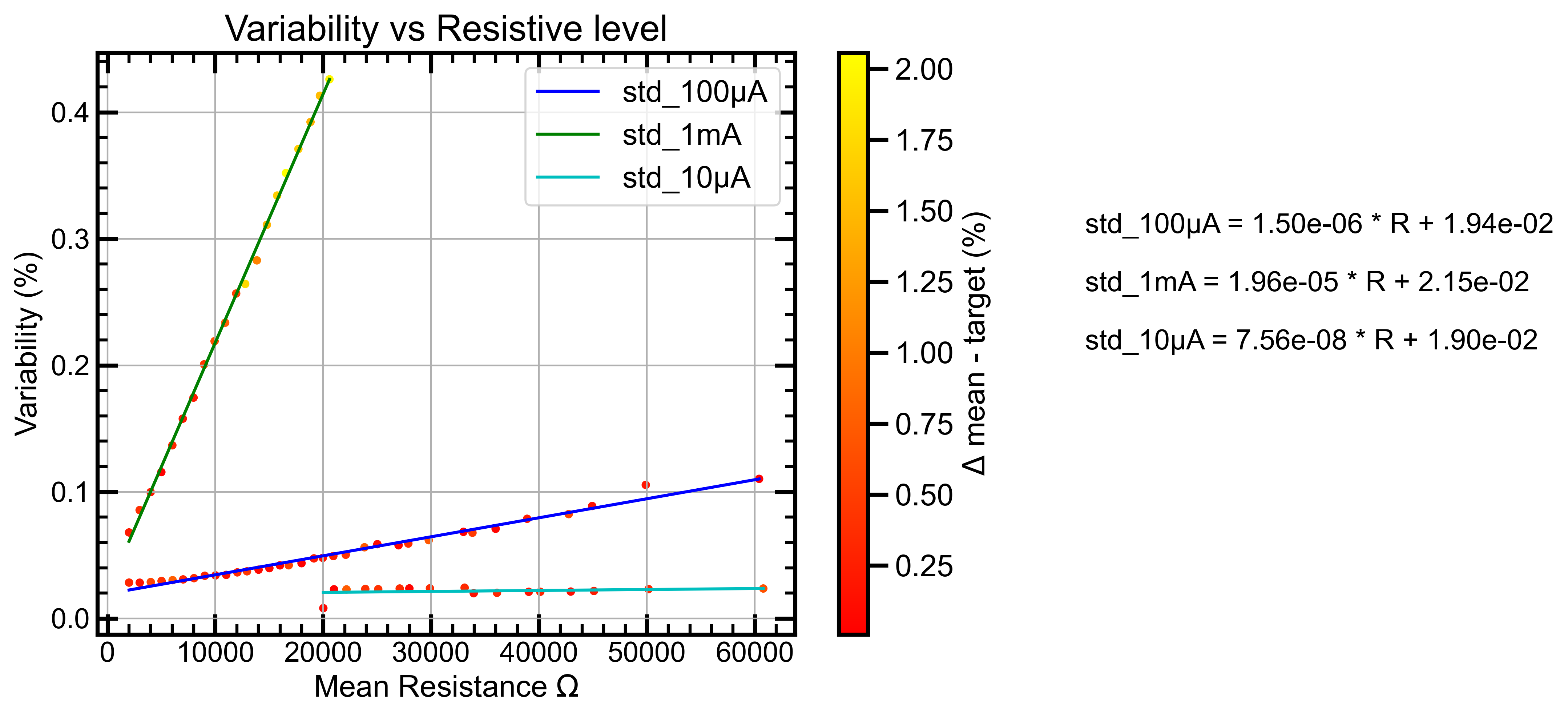}
    \caption{\textbf{B1500 read variability.} Several through-hole resistances were read 1e6 times between 1k $\Omega$ and 60k $\Omega$ to measure the inherent variability of the instrument with respect to the resistive value. The y value represents the standard deviation of the read values (as a percentage) and the heatmap on the right represents the drift of the mean of the measured values from the actual value of the resistor (note that the manufacturer guaranteed a maximal offset of 0.1\% of the manufactured resistances from their target values).  The variability is expressed as a percentage of the mean. The three differently colored lines represent the measurement ranges of the instrument. For the studied range of conductances, the variability never reaches more than 0.05\% of the mean value for the 100 $\mu$A current range used.}
    \label{fig:b1500_varia}
\end{figure*}

\end{document}